\documentclass[11pt]{article}
\usepackage{latexsym,amsmath,amsthm,amssymb}
\usepackage{graphicx}
\usepackage{subfigure}
\renewcommand{\baselinestretch}{1.5}
\newcommand{\mysum}{\displaystyle\sum}
\newcommand{\myprod}{\displaystyle\prod}

\newcommand{\bs}{\boldsymbol}
\newcommand{\xVec}{\mathbf{x}}
\newcommand{\tVec}{\mathbf{t}}
\newcommand{\thetaVec}{\bs{\theta}}
\newcommand{\etal}{et al.}
\DeclareSymbolFont{AMSb}{U}{msb}{m}{n}
\DeclareMathSymbol{\N}{\mathbin}{AMSb}{"4E}
\DeclareMathSymbol{\Z}{\mathbin}{AMSb}{"5A}
\DeclareMathSymbol{\R}{\mathbin}{AMSb}{"52}
\DeclareMathSymbol{\Q}{\mathbin}{AMSb}{"51}
\DeclareMathSymbol{\I}{\mathbin}{AMSb}{"49}
\DeclareMathSymbol{\C}{\mathbin}{AMSb}{"43}
\DeclareMathSymbol{\D}{\mathbin}{AMSb}{"44}
\DeclareMathSymbol{\J}{\mathbin}{AMSb}{"4A}

\title{Prediction and Computer Model Calibration\\[-7pt] Using Outputs From Multi-fidelity Simulators}
\author{Joslin Goh and Derek Bingham\\[-7pt]
        Department of Statistics and Actuarial Science\\[-7pt] Simon Fraser University \\[-7pt]
        Burnaby, BC, V5A 1S6, Canada\\[-7pt]
        \\
        James Paul Holloway, Michael J. Grosskopf \\[-7pt] Carolyn C. Kuranz and Erica Rutter\\[-7pt]
        Center for Radiative Shock Hydrodynamics \\[-7pt]
        University of Michigan \\[-7pt]
        Ann Arbor, MI, 48109, USA\\[-7pt]
 }
\date{}
\begin{document}
\maketitle

\renewcommand{\baselinestretch}{1.0}
\begin{abstract} 
Computer codes are widely used to describe physical processes in lieu of physical observations.  In some cases, more than one computer simulator, each with different degrees of fidelity, can be used to explore the physical system.  In this work, we combine field observations and model runs from deterministic multi-fidelity computer simulators to build a predictive model for the real process.  The resulting model can be used to perform sensitivity analysis for the system, solve inverse problems and make predictions.  Our approach is Bayesian and will be illustrated through a simple example, as well as a real application in predictive science at the Center for Radiative Shock Hydrodynamics at the University of Michigan. \\

\end{abstract}

\section{Introduction}
Deterministic computer models are used to simulate a wide variety of physical processes (Sacks \etal, 1989; Santner et al., 2003). Oftentimes, a single run of the code requires considerable computational effort, making it infeasible to continually exercise the simulator.  Instead, experimenters attempt to explore the computer model response (and to some extent the physical process) using a limited number of computer model runs. 

In some applications, several simulators of the physical process are available to describe the same system (Craig \etal, 1998; Kennedy and O'Hagan, 2000; Qian \etal, 2006 and 2008; Reese \etal, 2004; Cumming and Goldstein, 2009), each with different levels of fidelity. The varying levels of fidelity can occur, for example, because of the presence of reduced order physics in lower fidelity models, different levels of accuracy specified for numerical solvers or solutions obtained on finer grids.  In these cases, a higher fidelity model is thought to better represent the physical process than a lower fidelity model, but also takes more computer time to produce an output than a lower fidelity model.  So, combining relatively cheap lower fidelity model runs with more costly high fidelity runs to emulate the high fidelity model has been an significant problem of interest (Kennedy and O'Hagan, 2000; Qian \etal, 2006 and 2008).   


Another important application of computer models is that of {\it calibration} (e.g., Kennedy and O'Hagan, 2001; Higdon \etal, 2004) where the aim is to combine simulator outputs with physical observations to build a predictive model and also estimate unknown parameters that govern the behaviour of the computer model.  The latter endeavour amounts to solving a sort of inverse problem, while the former activity is a type of regression problem.  

Motivated by applications at the Center for Radiative Shock Hydrodynamics (CRASH) at the University of Michigan,  the aim of this work is to develop new methodology to combine outputs from simulators with different levels of fidelity and field observations to make predictions of the physical system with associated measurements of uncertainty.  In the spirit similar to Kennedy and O'Hagan (2000 and 2001) and Higdon \etal  ~(2004), 
we propose a predictive model that incorporates computer model outputs and field data, while attempting to find optimal values for some input parameters (i.e. \textit{calibration parameters}).  Different models are specified for each source of data (Kennedy and O'Hagan, 2000; Qian \etal, 2006 and 2008).  The approach calibrates each computer model to the next highest level of fidelity model, and the simulator of the highest fidelity is then calibrated to the field measurements.  All the response surfaces are Gaussian process (GP) models and the various sources of information that inform predictions of the physical system are combined with a Bayesian hierarchical model.   
  
The paper is organized as follows:  In section \ref{sec:method}, we will introduce the proposed methodology and the GP models involved, along with the relevant priors.  The framework for prediction will be discussed at the end of the section.  A simple example from the literature and an application from CRASH are used to demonstrate the proposed approach in Section \ref{sec:eg}.  Further discussion follows in Section \ref{sec:discussion} with some concluding remarks in Section \ref{sec:conclusion}.

\section{A Hierarchical Model for Multi-fidelity Model Calibration}\label{sec:method}
In this section, a Bayesian hierarchical model that calibrates multi-fidelity computer simulators is proposed.  The higher fidelity code is assumed to better represent the real world process but is assumed to require more computing resources to simulate the system.
For ease of presentation and notation, we present the case where there are only two computer simulators -- a high fidelity and a low fidelity model.   It is easy to extend the proposed methodology to cases with more than two simulators, and this setting is discussed in  Section \ref{sec:discussion}.

\subsection{The Hierarchical Model}\label{sec:model}
Throughout this work, the computer models are assumed to be deterministic mathematical functions that map inputs to outputs.  The computer codes have two types of inputs: (i) {\em design variables}, $\xVec$, that are adjustable or measurable in the field experiments; and (ii) {\em calibration parameters}, $\tVec$,  whose values are thought to impact the physical system, but are unknown a priori.  The latter inputs can only be adjusted within the simulator, but are not measurable in the field.  In model calibration problems, the issue is to build a predictive model for the field process and also to estimate the unknown calibration parameters. 

A unique feature of the application that motivated the current work is that the calibration parameters for the computer models are not all the same.  Some of the calibration parameters,  $\tVec_f$, are shared among the simulators, whereas others are required inputs only to individual simulators. The vectors of calibration inputs exclusive to the high and low fidelity models are denoted as $\tVec_h$ and $\tVec_l$, respectively.  

First consider the low fidelity computer model with inputs  $\left(\xVec, \tVec_f , \tVec_l \right)$ (i.e., the design variables and calibration parameters that are shared and unshared with the high fidelity simulator), where $\xVec= \left(x_1, \ldots , x_p\right)$, $\tVec_f= \left(t_{f,1},\ldots , t_{f,m_f}\right)$ and $\tVec_l= \left(t_{l,1},\ldots , t_{l,m_l}\right)$.  Outputs $Y_l(\cdot)$ from the low fidelity simulator, $\eta_l(\cdot)$, are written as:
\begin{equation}
 Y_l\left(\xVec, \tVec_{f} ,\tVec_{l}\right)=\eta_l\left(\xVec, \tVec_{f} ,\tVec_{l}\right).
\label{eqn:LFsim}\end{equation}
Similarly, the high fidelity simulator, $\eta_h(\cdot)$, has inputs $\left(\xVec, \tVec_f , \tVec_h \right)$, where   $\tVec_h = \left(t_{h,1},\ldots , t_{h,m_h}\right)$, and output $Y_h(\cdot)$:
\begin{equation*}
Y_h (\xVec , \tVec_{f}, \tVec_{h}) = \eta_h(\xVec, \tVec_{f}, \tVec_{h}).
\label{eqn:HFsim}\end{equation*}

Both simulators are used to describe the same physical process, but will not always give the same response.  There are a few obvious reasons why this is the case.   The lower fidelity model is inferior to the high fidelity simulator since it may, for example, fail to capture some processes that the high fidelity code can more accurately model.  Furthermore, the two codes do not share all of the same inputs.  The inputs $\tVec_{h}$ only appear in the high fidelity model and thus, any impact that these variables have on the output cannot be captured by the low fidelity model.  Similarly, the inputs $\tVec_{l}$ appear only in the low fidelity model. To address these issues, we  take the approach of writing the high fidelity simulator as a discrepancy adjusted version of the low fidelity model (e.g. Kennedy and O'Hagan, 2000; Qian \etal, 2006):
%
%
\begin{equation}
Y_h (\xVec, \tVec_{f}, \tVec_{h}) = \eta_l(\xVec, \tVec_f , \thetaVec_l) +\delta_2(\xVec, \tVec_{f} , \tVec_{h}).
\label{eqn:sim2withSim1}\end{equation}

Specifying the first term  in (\ref{eqn:sim2withSim1}) as $\eta_l(\xVec, \tVec_f , \thetaVec_l)$ amounts to partially calibrating (partially in the sense that the other calibration parameters must still be estimated) the first computer model to the second computer model.  Furthermore, the discrepancy, $\delta_2(\cdot)$, represents the systematic differences between the partially calibrated low fidelity model and the high fidelity code.  Lastly, notice that $\delta_2(\cdot)$ is a function of not only the design variables - as in Kennedy and O'Hagan (2001) - but also $(\tVec_f, \tVec_h)$.  The calibration parameters are included in this discrepancy term because they can be modified in the high fidelity code.  Therefore, this discrepancy term captures the systematic differences in the outputs from the two models over values of the design variables  and the changes in the calibration inputs $\tVec_f $ and $\tVec_h $.

In addition to the simulations, there are also field observations that are used to inform predictions.  Since the higher fidelity simulator is assumed to better represent the physical process than the low fidelity simulator, it is natural to model the field observations with the simulator of higher fidelity.  Similar to Kennedy and O'Hagan (2001), a discrepancy function, $\delta_f(\cdot)$, is used to capture the systematic inadequacy of the high fidelity simulator. The field observations are noisy versions of the mean process, and independent and identically distributed (iid) observational errors are included in our specification.  For input setting, $\xVec$, the field process is written as: 
\begin{equation}
Y_f(\xVec) = \eta_h(\xVec, \thetaVec_f, \thetaVec_h) + \delta_f(\xVec) +\epsilon, 
\label{eqn:fieldSim2}\end{equation}
where $\epsilon \sim N(0,1/\lambda_y)$.
Substituting  (\ref{eqn:sim2withSim1}) into (\ref{eqn:fieldSim2}) allows the field observations to be written:
\begin{equation}
Y_f(\xVec) = \eta_l(\xVec, \thetaVec_f , \thetaVec_l) +\delta_2(\xVec, \thetaVec_f , \thetaVec_h) + \delta_f(\xVec) +\epsilon,
\label{eqn:fieldFinal}\end{equation}

So, the response surface for the field data is written as the sum of the calibrated low fidelity simulator,  the calibrated discrepancy between the two different simulators, the discrepancy between the high fidelity model and the data, and observational error.  From here on out, we describe the response surfaces for the low and high fidelity simulators and the field data using the framework described in equations (\ref{eqn:LFsim}), (\ref{eqn:sim2withSim1}) and (\ref{eqn:fieldFinal}), respectively.

It is possible at this point to envision applications with more than two simulators, each ranked from low to high by levels of fidelity.  The above framework can be extended to these cases.  For example, in the case where there are three simulators of different fidelity, $\eta_1(\cdot)$, $\eta_2(\cdot)$ and $\eta_3(\cdot)$, where $\eta_1(\cdot)$ is of the lowest fidelity and $\eta_3(\cdot)$ is best at describing the physical process.  Outputs from $\eta_1(\cdot)$ and $\eta_2(\cdot)$ are modelled as above.  Similarly, $\eta_3(\cdot)$ will be modelled using $\eta_1(\cdot)$ and some discrepancy functions, and the field observations will be modelled with the highest fidelity simulator $\eta_3(\cdot)$.  More on this in Section \ref{sec:discussion}.

\subsection{Gaussian Process Models}\label{sec:GPprior}
To make predictions of the physical system, response surfaces for the computer model and discrepancies must be estimated.   We follow the common practice of emulating  simulator responses using a GP (e.g., see Sacks et al., 1989). The reason for this, in general, boils down to the success of the GP as a non-parametric regression estimator and also the ability of the GP model to provide a basis for statistical inference for the outputs of deterministic computer codes.  From a Bayesian viewpoint in this context, one can think of the GP as a prior distribution over the class of functions produced by the low fidelity simulator and the discrepancies, respectively.

We begin by first considering the specification for the low fidelity simulator.  The outputs are treated as a realization of a random function of the form:
\begin{equation*}\label{eqn:lfgp}
Y_l\left(\xVec, \tVec_{f} ,\tVec_{l}\right)= \sum_{i=1}^p f_i \left( \xVec, \tVec_{f} ,\tVec_{l}\right)\beta_i + Z \left( \xVec, \tVec_{f} ,\tVec_{l} \right),
\end{equation*}
where $f_1, \ldots, f_p$ are regression functions, ${\boldsymbol \beta} = (\beta_1, \ldots, \beta_p)'$ is the vector of unknown regression coefficients, and $Z$ is a mean zero GP. We follow the convention of most computer model applications by specifying the mean function as a constant, $\mu$, and model the response surface through the covariance structure.   The covariance between observations at inputs  $\left(\xVec_i ,\tVec_{f,i}, \tVec_{l,i}\right)$ and $ \left(\xVec_j ,\tVec_{f,j}, \tVec_{l,j}\right)$ is specified as
\begin{eqnarray}
Cov\left[Z\left(\xVec_i ,\tVec_{f,i}, \tVec_{l,i}\right), Z\left(\xVec_j ,\tVec_{f,j}, \tVec_{l,j}\right)\right] &=& \frac{1}{\lambda_{\eta_l}} \myprod_{s=1}^{p}\rho_{\eta_l,s}^{4\left(x_{i,s}-x_{j,s}\right)^2} \myprod_{s=1}^{m_f} \rho_{\eta_l,p+s}^{4\left(t_{f,i,s}-t_{f,j,s}\right)^2}\myprod_{s=1}^{m_l} \rho_{\eta_l,p+m_f+s}^{4\left(t_{l,i,s}-t_{l,j,s}\right)^2}\nonumber\\
&=& \frac{1}{\lambda_{\eta_l}} R\left(\left(\xVec_i ,\tVec_{f,i},\tVec_{l,i}\right), \left(\xVec_j ,\tVec_{f,j},\tVec_{l,j}\right),\bs\rho_{\eta_l}\right),
\label{eqn:eta1Cov}\end{eqnarray}
\noindent where  $\lambda_{\eta_l}$ is the marginal precision of the simulator $\eta_l$. The $\left(p+m_f+m_l\right)$-vector $\bs\rho_{\eta_l} = (\rho_{\eta_l,1},\ldots,$ $ \rho_{\eta_l,p}, \ldots ,\rho_{\eta_l,p+m_f},\ldots ,\rho_{\eta_l,p+m_f+m_l})$ is the vector of correlation parameters that govern the dependence in each of the component directions of $\xVec$, $\tVec_f$ and $\tVec_l$  (e.g., Sacks et al., 1989; Higdon \etal, 2004 and 2008; Linkletter \etal, 2006).  

The discrepancy,  $\delta_2(\cdot)$, captures the systematic differences between the high and low fidelity simulators 
as a function of the inputs, ($\xVec, \tVec_h ,\tVec_f $), that are adjustable in the high fidelity model.  Continuing as above,  $\delta_2(\cdot)$ is modelled as mean zero GP with covariance: 

\begin{eqnarray}
Cov\left[Z\left(\xVec_i,\tVec_{f,i},\tVec_{h,i}\right), Z\left(\xVec_j, \tVec_{f,i},\tVec_{h,j}\right)\right] &=& \frac{1}{\lambda_2}  \myprod_{s=1} ^p \rho_{2,s}^{4\left(x_{i,s}-x_{j,s}\right)^2}\myprod_{s=1}^{m_f} \rho_{2,p+s}^{4\left(t_{f,i,s}-t_{f,j,s}\right)^2} \myprod_{s=1} ^{m_h} \rho_{2,p+m_f+s}^{4\left(t_{h,i,s}-t_{h,j,s}\right)^2}\nonumber\\
&=&\frac{1}{\lambda_{2}} R\left(\left(\xVec_i ,\tVec_{f,i},\tVec_{h,i}\right), \left(\xVec_j ,\tVec_{f,j},\tVec_{h,j}\right),\bs\rho_{2}\right),
\label{eqn:deltaCCov}\end{eqnarray}
where the marginal precision of the discrepancy function is $\lambda_2$, and the vector of correlation parameters for the inputs is $\bs\rho_2 = \left (\rho_{2,1},\ldots ,\rho_{2,p}, \ldots, \rho_{2,p+m_f},\ldots , \rho_{2,p+m_f+m_h}\right )$.

The  function $\delta_f(\cdot)$ is the discrepancy between the response from high fidelity simulator and the physical process.  Again, a zero mean GP is chosen.  Let $\lambda_f$ denote the marginal precision of the discrepancy function, $\delta_f(\cdot)$, and the vector $\bs\rho_f=\left(\rho_{f,1},\cdots \rho_{f,p}\right)$ be the correlation parameters for the $p$ design variables.  The covariance function is written as

\begin{eqnarray}
Cov\left[Z(\xVec_i) ,Z(\xVec_j)\right] &=& \frac{1}{\lambda_f} \myprod_{s=1}^p \rho_{f,s}^{4\left(x_{i,s}-x_{j,s}\right)^2}\nonumber\\
&=&\frac{1}{\lambda_f} R\left(\left(\xVec_i , \xVec_j\right),\bs\rho_f \right).
\label{eqn:deltaFCov}\end{eqnarray}


We define the vector of all observations and simulation outputs as  $\mathbf{Y}=\left(\mathbf{Y}^T_f,\mathbf{Y}^T_h,\mathbf{Y}^T_l\right)^T$, where $\mathbf{Y}_f =\left(Y_f(\xVec_1),\ldots ,Y_f(\xVec_{n_f})\right)^T$ is the vector of $n_f$ field measurements, and the vectors $n_h$ and $n_l$ simulated runs from the high and low fidelity simulators are $\mathbf{Y}_h =\left(Y_h(\xVec^{'}_1,\tVec^{'}_{f,1},\tVec^{'}_{h,1}),\ldots ,Y_h(\xVec^{'}_{n_h},\tVec^{'}_{f,n_h},\tVec^{'}_{h,n_h})\right)^T$ and  $\mathbf{Y}_l =\left(Y_l(\xVec^{*}_1,\tVec^{*}_{f,1},\tVec^{*}_{l,1}),\ldots ,Y_l(\xVec^{*}_{n_l},\tVec^{*}_{f,n_l},\tVec^{*}_{l,n_l})\right)^T$, respectively.  To simplify notation, denote $\bs\theta = \left(\bs\theta_f,\bs\theta_h,\bs\theta_l\right)$, $\bs\lambda = \left(\lambda_{\eta_l},\lambda_2,\lambda_f\right)$ and $\bs\rho = \left(\bs\rho_{\eta_l},\bs\rho_2,\bs\rho_f\right)$.  The likelihood for $\mathbf{Y}$ is

\begin{eqnarray*}
L\left(\mathbf{Y} |\bs\theta, \bs\mu , \bs\lambda , \bs\rho \right) \propto \left | \Sigma_{\mathbf{Y}}\right |^{-\frac{1}{2}}\exp\left\{\left(\mathbf{Y}-\bs\mu\right)^T\Sigma_{\mathbf{Y}}^{-1}\left(\mathbf{Y}-\bs\mu\right)\right\},
\label{eqn:zLikeliFunc}\end{eqnarray*}

\noindent  where $\bs\mu$ is the constant mean vector and 

\begin{equation}
\Sigma_{\mathbf{Y}} = \Sigma_{\eta_l} + \left (\begin{array}{cc}
\Sigma_2 & 0 \\
 0 & 0 \end{array} \right)
 + \left (\begin{array}{ccc}
\Sigma_f+ \Sigma_y& 0 & 0 \\
 0 & 0 & 0 \\
 0 & 0 & 0  \end{array} \right).
\label{eqn:zCovMat}\end{equation}

\noindent The covariance matrix $\Sigma_{\eta_l} $ is the covariance between all the outputs and is obtained by applying equation (\ref{eqn:eta1Cov}) to each pair of the $(n_f+n_h+n_l)$ input settings of the observed and simulated data.  Similarly, the covariance matrix $\Sigma_2 $ describes the relationship of the systematic difference between the two simulators and hence, is obtained by applying equation (\ref{eqn:deltaCCov}) to each pair of the $n_f$ input settings of the observed data $\mathbf{Y}_f$ and $n_h$ input settings of the simulated data $\mathbf{Y}_h$. It is a square matrix of order $(n_f+n_h)$.  Equation (\ref{eqn:deltaFCov}) is only applied to each pair of the $n_f$ input settings for the covariance matrix $\Sigma_f$.  
The covariance matrix for the measurement error ($\epsilon$) is given by the $n_f \times n_f$ diagonal matrix $\Sigma_y = (1/ \lambda_y) I_{n_f}$ and $\lambda_y$ is the precision parameter of the observational error.

\subsubsection{Priors for the GPs and MCMC}\label{sec:hyperPrior}
The posterior distribution of calibration and statistical model parameters, $\left(\bs\theta, \mu , \bs\lambda , \bs\rho \right)$, takes the form

\begin{equation}
\bs\pi \left( \bs\theta, \bs\mu , \bs\lambda , \bs\rho | \mathbf{Y}\right) \propto L\left(\mathbf{Y} |\bs\theta, \bs\mu , \bs\lambda , \bs\rho \right) \times \pi\left(\bs\theta\right)\times\pi\left(\mu\right) \times\pi\left(\bs\lambda\right) \times\pi\left(\bs\rho\right),
\label{eqn:zPostDensity}\end{equation}
where we abuse notation for prior distributions and denote the prior for $\bs\theta$, $\bs\lambda$ and $\bs\rho$ as $\pi\left(\bs\theta\right)=\myprod_{i=1}^{m_f}\pi\left(\theta_{f,i}\right) \times \myprod_{i=1}^{m_h}\pi\left(\theta_{h,i}\right)\times \myprod_{i=1}^{m_l}\pi\left(\theta_{l,i}\right)$, $\pi\left(\bs\lambda\right) = \pi\left(\lambda_{\eta_l}\right) \times \pi\left(\lambda_2\right) \times \pi\left(\lambda_f\right)\times\pi\left(\lambda_y\right)$ and $\pi\left(\bs\rho\right) = \myprod_{i=1}^{p+m_f+m_l}\pi\left(\rho_{\eta_{l,i}}\right)\times \myprod_{i=1}^{p+m_f+m_h} \pi\left(\rho_{2,i}\right)\times \myprod_{i=1}^p \pi\left(\rho_{f,i}\right)$, respectively.\\

In practice, we have to estimate $\bs\theta$, $\bs\lambda$ and $\bs\rho$.  We use Markov chain Monte Carlo (MCMC) to sample from the posterior distribution of the parameters, given observations and simulations.  
In order to simplify the prior specification, the responses are standardized to have mean zero and variance 1 (e.g., Linkletter \etal, 2006).  Hence, the prior for the precision of the marginal variance, $\lambda_{\eta_l}$, is chosen to encourage its values to be close to $1$ - the idea being that the low fidelity model should capture much of the signal in the observations.  We use a Gamma distribution for the prior for $\lambda_{\eta_l}$:

\begin{equation*}
\pi\left(\lambda_{\eta_l}\right) \propto \lambda_{\eta_l} ^{a_{\eta_l}}\exp\left\{-b_{\eta_l}\lambda_{\eta_l}\right\}.
\label{eqn:gammaPrior}\end{equation*}

\noindent When expert knowledge is unavailable,  we have found that $a_{\eta_l}=b_{\eta_l}=5$ (Higdon \etal , 2004) works reasonably well as the choice centers the prior distribution at 1 with a reasonably large variance, thereby allowing for a fairly broad exploration of the posterior.  Similarly, the priors chosen for the remaining precision parameters are also Gamma distributions. 
We also use the default values proposed by  Higdon et al. (2004) for the hyperparameters of priors for the remaining precision parameters.  The default choice of shape and scale parameters are $a^*=1$ and $b^*=0.001$.  This specification implies a relatively uninformative prior for these precision parameters, which encourages the data to choose a suitable value by itself.
  
The components in $\bs\rho$ are bounded within the unit interval.  Hence, a natural choice of prior for $\rho^*\in\bs\rho$ is the Beta distribution of the following form:

\[
\ \pi\left(\rho^*\right) \propto (\rho^*)^{a^*-1} \left(1-\rho^*\right)^{b^*-1}.
\]
Conventionally, the Beta priors are flat and centered at 1 with small variance (Williams \etal, 2006).  This is based on the prior belief that all the inputs are equally uncorrelated to the simulator and allow the data to decide the dependence of the simulator on the different inputs by moving the $\rho$'s away from 1 in the posterior.  In our experience, the default choices for these parameters, $a^*= 1$ and $b^*=0.001$, suggested by Higdon et al. (2004) and Williams et al. (2006) encourage strong enough dependence in each of the parameters and work well in general. 

The posterior distribution for each parameter is explored using MCMC. Specifically, single-site Metropolis updates (Metropolis et al., 1953) are used for the components of $\bs\rho$ and $\bs\theta$. Proposals are made for each of these parameters from a uniform distribution centered at the parameter's current value.  The widths of the uniform distributions (one for each component parameter) are pre-computed by running short MCMC runs and choosing a width that gives an acceptance rate of about 0.44 (Gelman et al., 2004).  Although this adjustment does not guarantee an acceptance probability of 0.44, we have found this procedure is helpful at choosing widths resulting in acceptance ratios between 0.25 and 0.75 and, more importantly, encourages the MCMC to converge.  Good default choices for the widths for the updates can also be found using the method proposed by Graves (2005).
For each of the precision parameters,  we used Hastings updates (Hastings, 1970), where the proposed value is drawn from a uniform distribution centered at the current parameter value, with a width that is proportional to the parameter's current value.  We have found that a width that is $0.3$ times the current parameter value (proposed by Higdon et al., 2008) works fairly well in general.  

\subsection{Prediction}
The main goal of this endeavour is prediction.  
Given the posterior realizations from (\ref{eqn:zPostDensity}), predictions of the field measurement, $Y_f(\xVec^{new})$, can be made at a new input setting $\xVec^{new}$.  
The joint distribution between $\mathbf{Y}$ and $Y_f(\xVec^{new})$, conditional on the parameters  $\thetaVec$, $\bs\lambda$ and $\bs\rho$,  is: 
\begin{eqnarray*}
\left( \begin{array}{c}
\mathbf{Y}\\
Y_f(\xVec^{new})
\end{array}\right) \sim MVN\left(\mathbf{0},\Sigma^{new}\right), 
\label{eqn:predDist}\end{eqnarray*}
\noindent where the covariance matrix, $\Sigma^{new}$, is analogous to the covariance in (\ref{eqn:zCovMat}) - there is an extra row and column in $\Sigma^{new}$ as a result of appending $Y_f(\xVec^{new})$ to  $\mathbf{Y}$.

Through the usual properties of the multivariate normal distribution, the predictive distribution of $Y_f(\xVec^{new})$, conditional on the parameters, is:
\begin{equation}
Y_f(\xVec^{new})\mid \mathbf{Y}\sim MVN\left(\mu_{pred},\Sigma_{pred}\right),
\label{eqn:ypred}\end{equation}
where
$\mu_{pred} = \Sigma_{21}^{new}\left(\Sigma_{11}^{new}\right)^{-1}\mathbf{Y}$ and $\Sigma_{pred} = \Sigma_{22}^{new}-\Sigma_{21}^{new}\left(\Sigma_{11}^{new}\right)^{-1}\Sigma_{12}^{new}$.  The matrices $\Sigma_{ij}^{new}$ are sub-matrices of $\Sigma^{new}$ where
\[
\ \Sigma^{new}= \left(\begin{array}{cc}
\Sigma_{11}^{new} & \Sigma_{12}^{new}\\
\Sigma_{21}^{new} & \Sigma_{22}^{new}\\
\end{array} \right).
\]The sub-matrix $\Sigma_{11}^{new}$ is an $\left(n_f+n_h+n_l\right)\times \left(n_f+n_h+n_l\right)$ matrix, while $\Sigma_{12}^{new}$ and $\Sigma_{21}^{new}$ are of dimension $\left(n_f+n_h+n_l\right)\times 1$ and $1\times \left(n_f+n_h+n_l\right)$, respectively. The remaining sub-matrix, $\Sigma_{22}^{new}$, is a scalar.

To make predictions, we first sample a vector of parameters from (\ref{eqn:zPostDensity}).  Next, conditional on the sampled parameters, a prediction is sampled from (\ref{eqn:ypred}).  The sampling of parameters and predictions are repeated many times to provide estimated posterior quantities (e.g., posterior mean, variance or prediction intervals). 

\section{Examples}\label{sec:eg}
In this section, two examples are presented.  The first example is a simple computer model that is used to demonstrate the proposed approach.  After illustrating the implementation of our approach and some diagnostics to assess the adequacy of the model fit, a  small simulation study is carried out to investigate the predictive performance of the proposed methodology.
The second example is the application that motivated this work, and involves a radiative shock experiment conducted at CRASH.  The main goal is to predict the observed field measurements given the outputs from two computer models and some field trials.

\subsection{Toy Example}\label{sec:toyEg}
We begin with the ``toy" example in Bastos and O'Hagan (2009), with some slight alterations.  That is, the setting has been modified to accommodate two computer models and field experiments.  In addition, we refashion the computer model to include two design variables, a common calibration parameter and calibration parameters that exist in each computer model, respectively.  For simplicity, all the input settings and calibration parameters are chosen from the unit interval.

We specify the low fidelity model as:
\begin{eqnarray}
y_l\left(\xVec,t_f,t_l\right)&=&\eta_l\left(\xVec,t_f,t_l\right) \nonumber\\
&=& \left(1-\exp(-\frac{1}{2x_2})\right)\frac{1000t_fx_1^3 + 1900x_1^2+2092x_1+60}{1000t_lx_1^3 + 500x_1^2+4x_1+20}.
\label{eqn:toyEta}\end{eqnarray}
The high fidelity model is defined as the low fidelity response model plus a discrepancy term:
\begin{eqnarray}
y_h\left(\xVec,t_f,t_h\right)&=&\eta_l\left(\xVec,t_f,\theta_l\right) + 5\exp(-t_f)\frac{x_1^{t_h}}{100\left(x_2^{2+t_h}+1\right)} \nonumber\\
&=&\eta_l\left(\xVec,t_f,\theta_l\right) + \delta_c\left(\xVec,t_f,t_h\right).
\label{eqn:toyDeltaC}\end{eqnarray}

To illustrate the proposed approach, we simulate outputs from the respective models.  Following Loeppky et al. (2009), we used a 40 run random Latin hypercube design (Mackay \etal, 1979) for the low fidelity simulator.  Since, in practice, the high fidelity model is likely to be more computationally expensive than the low fidelity model, only 10 runs  are generated -- also chosen using a random Latin hypercube design.   

In most computer model applications, there are relatively few field observations.  Consequently to mimic this setting, only 3 field observations were simulated from the mathematical model:
\begin{eqnarray}
y_f\left(\xVec\right) &=& \eta_l\left(\xVec,\theta_f,\theta_l\right) +\delta_c\left(\xVec,\theta_f,\theta_h\right)+ \frac{10x_1^2+4x_2^2}{50x_1x_2+10} + \epsilon\nonumber\\
&=& \eta_l\left(\xVec,\theta_f,\theta_l\right) +\delta_c\left(\xVec,\theta_f,\theta_h\right)+\delta_f\left(\xVec\right)+\epsilon,
\label{eqn:toyDeltaF}\end{eqnarray}
where $\epsilon\sim N(0,0.5^2)$.   

In this set-up, the true value of the common calibration parameter is chosen to be $\theta_f = 0.2$, while the calibration parameter appearing only in the high and low fidelity models are chosen to be $\theta_h = 0.3$ and $\theta_l=0.1$, respectively.

\begin{figure}[h!]
\centering
\includegraphics[scale=0.8]{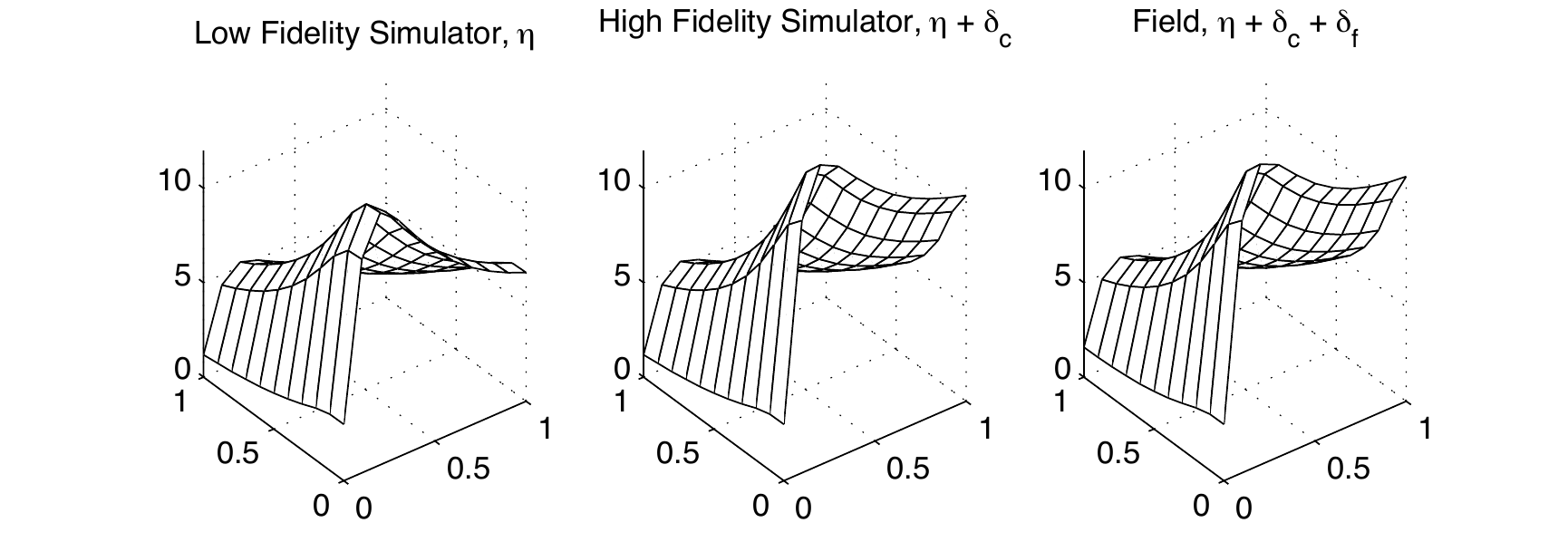}
\caption{From left to right: response surface of the low fidelity simulator, the high fidelity simulator and the mean of the physical process as outlined in equations (\ref{eqn:toyEta})-(\ref{eqn:toyDeltaF}).}
\label{fig:surfaces}\end{figure}

Figure \ref{fig:surfaces} displays the response surfaces for the two computer models and also the mean response surface for the field process.  A quick glance at the figure reveals that the high fidelity model appears closer to the mean process than the low fidelity model.  This represents the framework we are working within insofar as the high fidelity model is expected to be more like the true system than the low fidelity model.

The posterior distribution of the model parameters was sampled using MCMC as outlined in Section \ref{sec:hyperPrior}. The MCMC chain is initialized with $\theta_f=\theta_h=\theta_l=0.5$ (i.e., the centre of the input space), $\lambda_{\eta_l} =1 $, $\lambda_2 =\lambda_f =\lambda_y= 20$ and all the correlation parameters, $\bs\rho$ are chosen to be $0.1$ as we assume that the simulator and discrepancy terms are dependent on all the inputs.  Through visual inspection of the traceplots (not shown), we found that, for the data encountered in this example, convergence is achieved in the first 1,000 steps are so.  The MCMC was run for 10,000 steps, where the first 2,000 steps are treated as burn-in and discarded in further analysis. 

\begin{figure}[h!]
\centering
\includegraphics[scale=0.7]{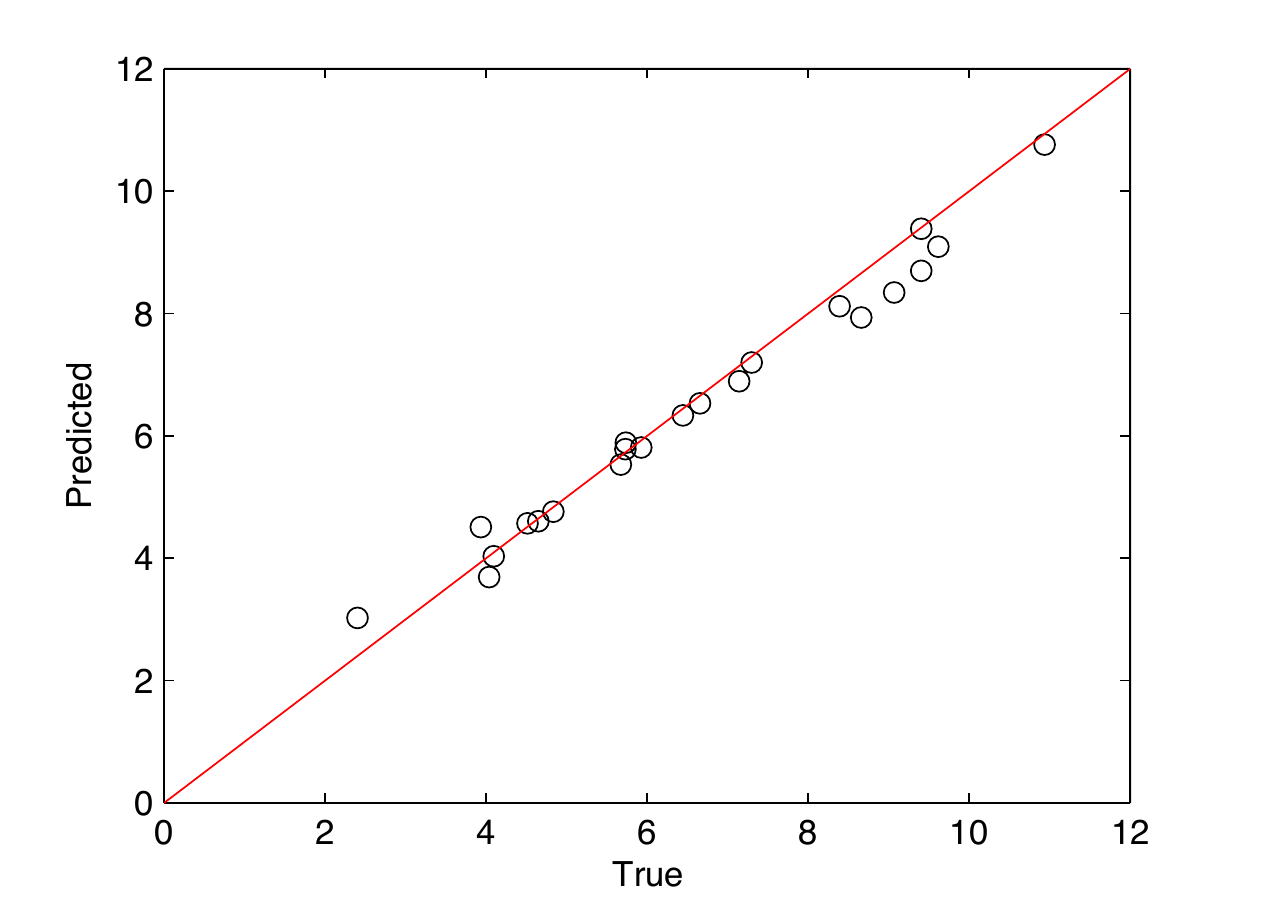}
\caption{Predicted versus actual field measurements of the validation set (with the $y=x$ line)}
\label{fig:samplePred}\end{figure}

In addition to the data simulated from (\ref{eqn:toyEta}) -- (\ref{eqn:toyDeltaF}) used to fit the proposed model (i.e., the training set), a validation dataset was generated from (\ref{eqn:toyDeltaF}), so that the predictive performance can be evaluated.  The validation set consisted of 25 field observations with input settings, $\xVec$, chosen using random Latin hypercube sampling.  
We use the posterior mean prediction at $\xVec$ to estimate $Y_f(\xVec)$.  Figure \ref{fig:samplePred} shows the predicted versus actual values for each of the validation points.  The figure shows that the predictive model performs reasonably well since the points center around the $y=x$ line.  

\begin{figure}[h!]
\centering
\includegraphics[scale=0.7]{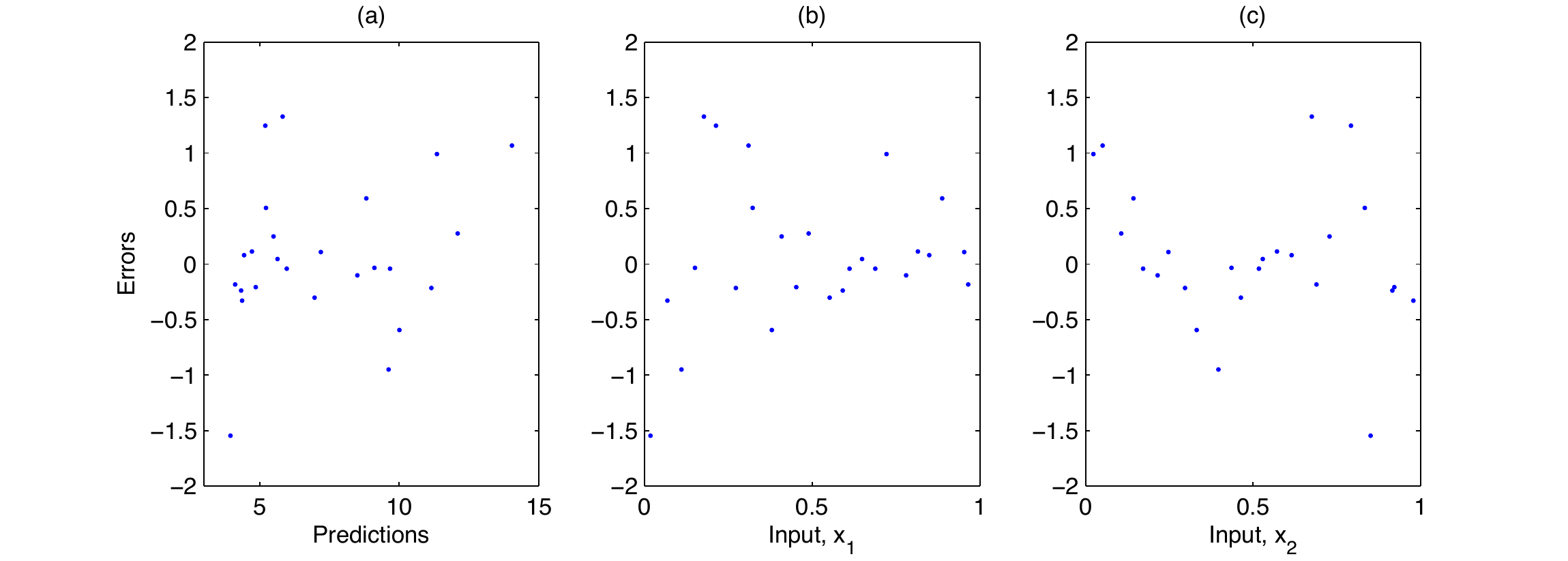}
\caption{Diagnostics plots for the simple example: (a) Prediction error against predictions; (b) Prediction error against $x_1$; (c) Prediction error against $x_2$. }
\label{fig:predVsStuff}\end{figure}

Figure \ref{fig:predVsStuff} displays the deviations of the predictions from the true values plotted against the predictions and also the input settings in each dimension.  In each case, no obvious pattern is found in the plots, suggesting the outputs have similar degree of smoothness across the input space and that no obvious systematic behaviour was unaccounted for.

\begin{figure}[h!]
\centering
\includegraphics[scale=0.6]{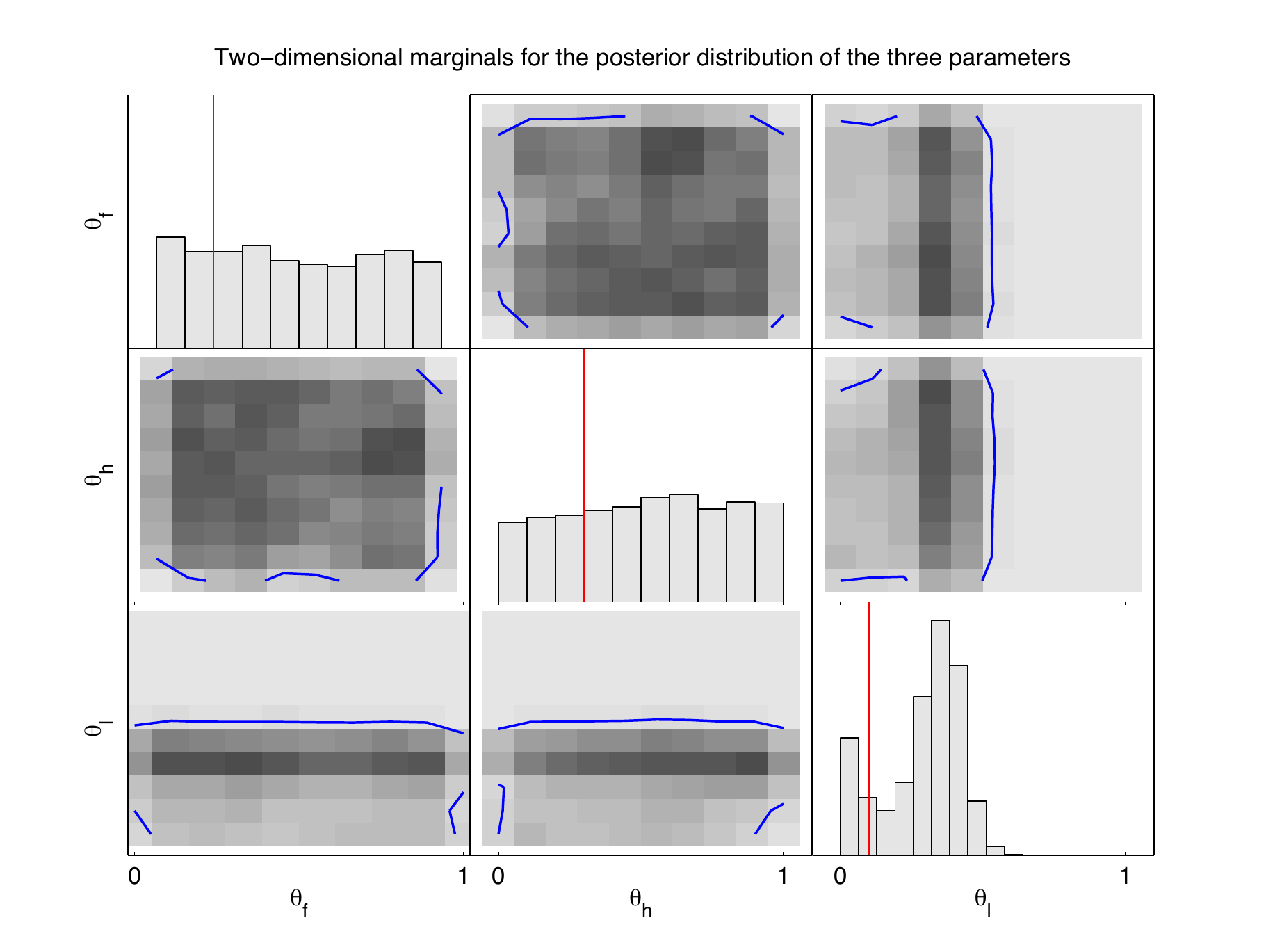}
\caption{The diagonals show the marginal posterior distributions of the calibration parameters, with the true values marked with vertical lines.  The off-diagonals sub-plots contain the two-dimensional marginal posterior distributions for the three calibration parameters.  The solid lines represent the 95\% high posterior density region.}
\label{fig:toyTheta}\end{figure}

While not the specific goal of the proposed methodology, we now consider the estimation of the calibration parameters.  Figure \ref{fig:toyTheta} shows the estimated one-dimensional and two-dimensional marginal posterior distributions of the calibration parameters.  Solid vertical lines are plotted at the true values of the calibration parameters.  In general, these posterior distributions can be interpreted as representing the uncertainty in the calibration parameters given the very limited number of observations and small numbers of simulations from imperfect computer models.  A quick glance at the plots reveals that, except for $\theta_l$, the calibration parameters are not being constrained by the data.  It is not too surprising that we can constrain $\theta_l$, but not the calibration other parameters, since there are more outputs (comparisons between the low and high fidelity models) to inform this parameter.  The inability to constrain the other calibration parameters is due to the presence of the discrepancy terms $\delta_f(\cdot)$, and the dearth of data.    

\begin{figure}[h!]
\centering
\subfigure[$n_l=n_h=20$, $n_f=3$]{
\includegraphics[scale=0.43]{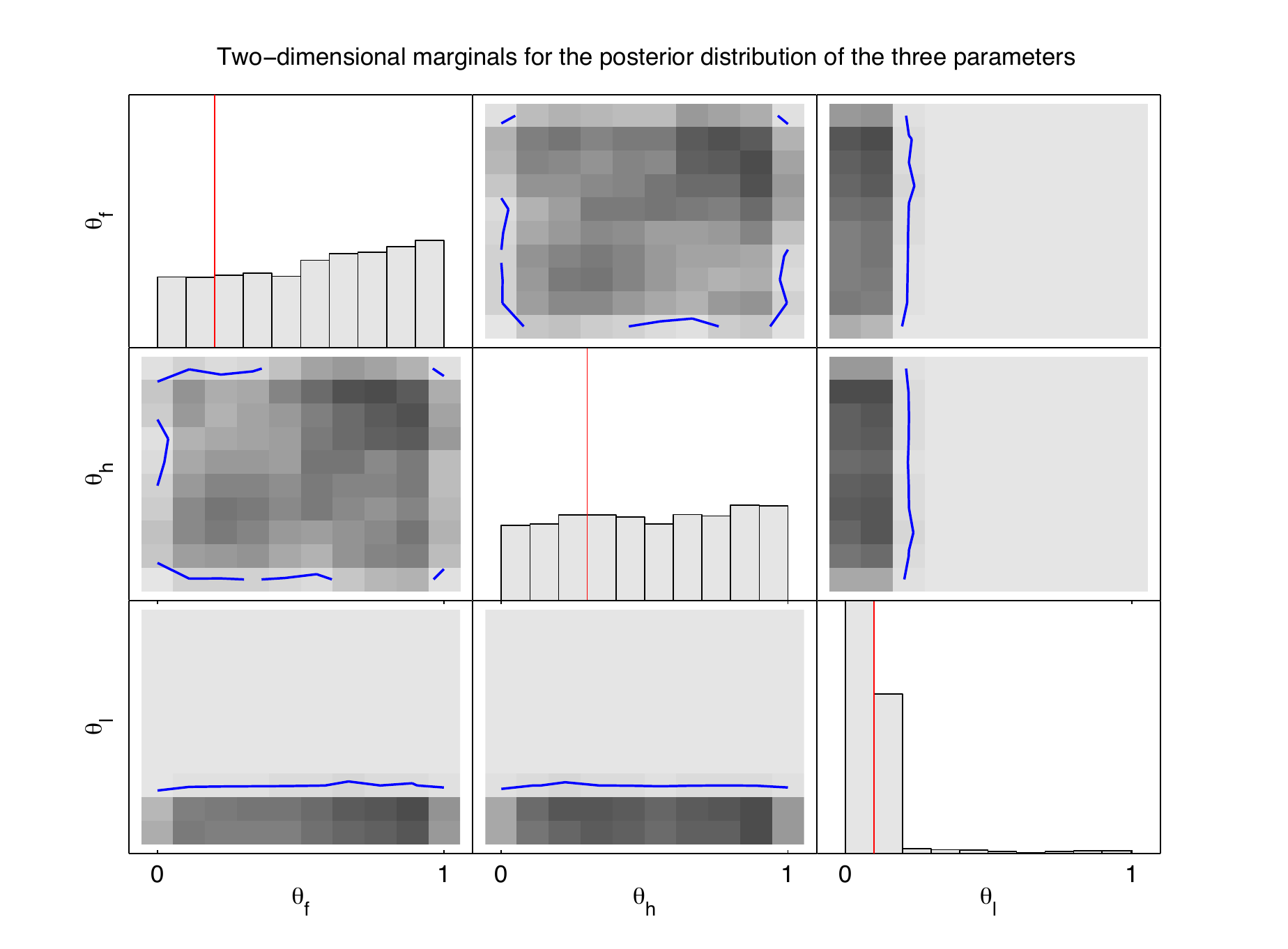}
   \label{fig:toy20part}
 }
 \subfigure[$n_l=n_h=n_f=40$]{
\includegraphics[scale=0.43]{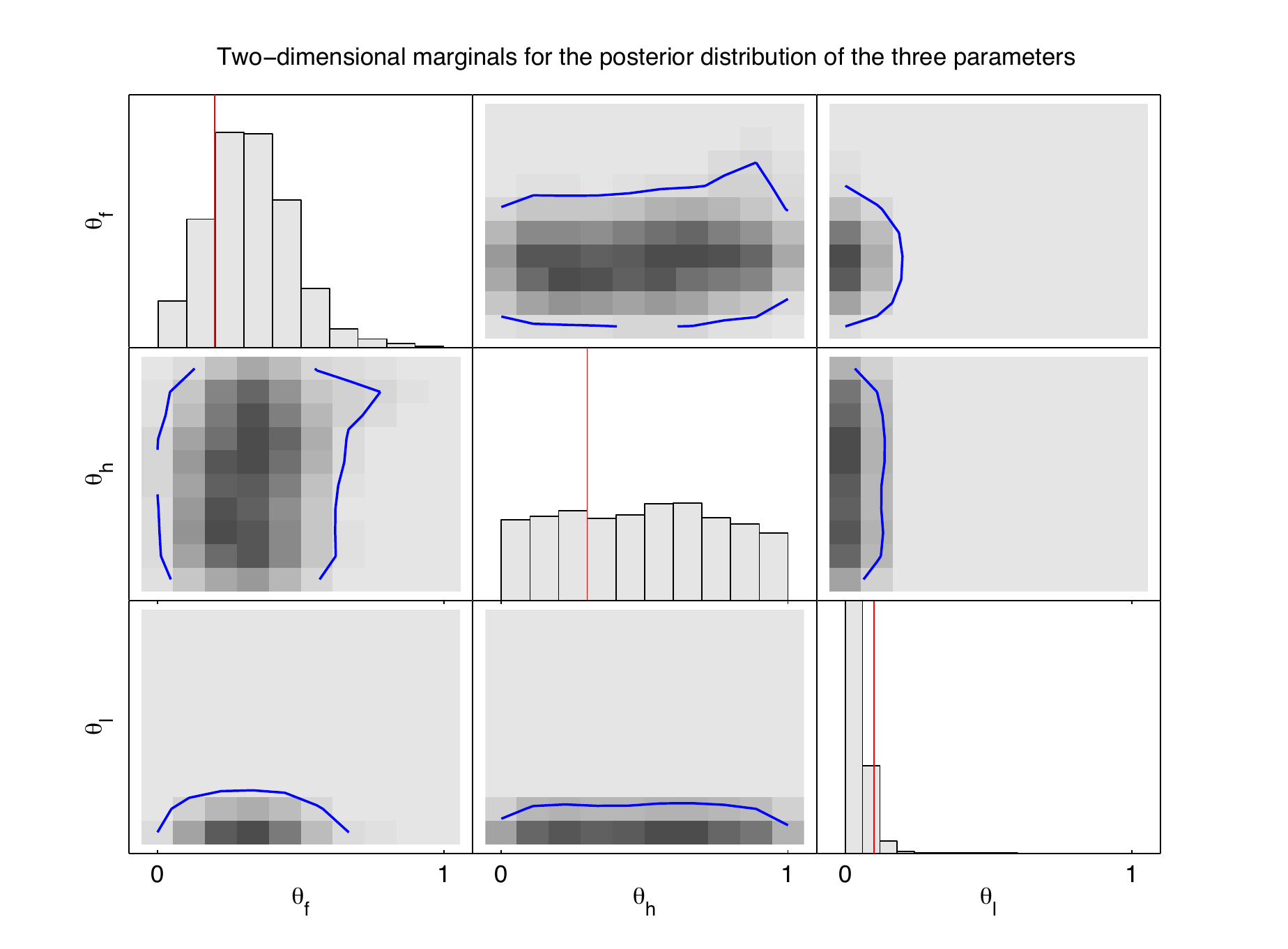}
   \label{fig:toy40}
 }
 \subfigure[$n_l=n_h=n_f=100$]{
   \includegraphics[scale =0.43] {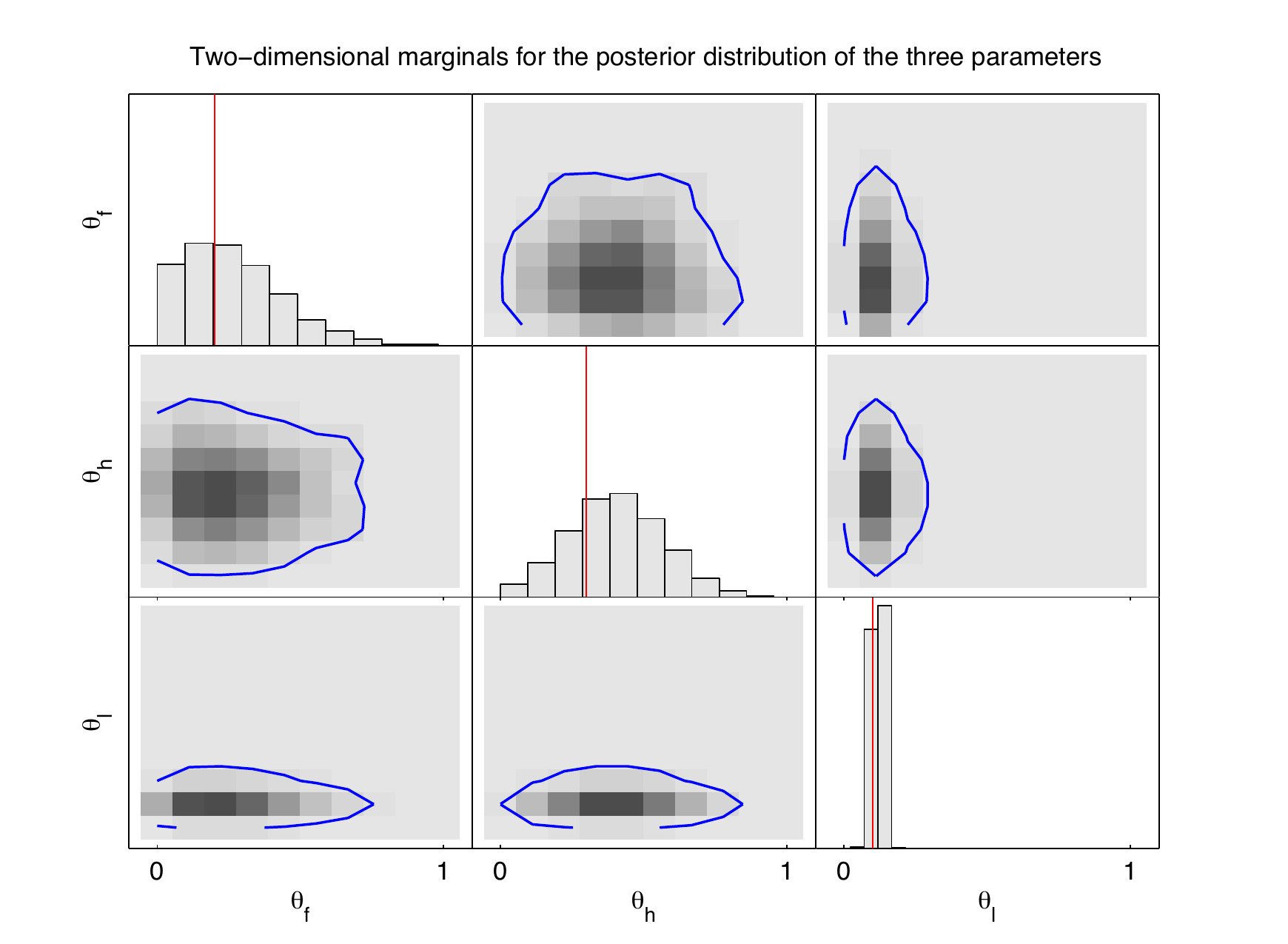}
   \label{fig:toy100}
 }
\caption{The diagonals show the marginal posterior distributions of the calibration parameters, with the true values marked with vertical lines.  The off-diagonals sub-plots contain the two-dimensional marginal posterior distributions for the three calibration parameters.  The solid lines represent the 95\% high posterior density region.}
\label{fig:toy40Theta}\end{figure}

Figure \ref{fig:toy40Theta} shows the estimated posterior distributions of the calibration parameters for different sample sizes.    Panels (a), (b) and (c)  are the results of the simulations with (a) $n_l=n_h=20$, $n_f=3$, (b) $n_l=n_h=n_f=40$ and  (c) $n_l=n_h=n_f=100$.  The first case was chosen as a more simulation rich version of the above example.  Comparing panel (a) of Figure  \ref{fig:toy40Theta} with the results in Figure \ref{fig:toyTheta}, we see that the mode of the posterior distribution of $\theta_l$ is closer to the true value (solid line) and there is less variability in the posterior distribution when there are more simulations.  However, very little is learned about the calibration parameters $\theta_h$ and $\theta_f$.  To gain more information on these parameters, there needs to be more field observations.  Panels (b) and (c) consider cases where the number of simulations and field trials is larger than before.  As the number of observations and simulations increases, the model is able to better estimate the calibration parameters. An interesting observation is that the shared calibration parameter $\theta_f$ is better constrained in panel (b) than $\theta_h$.  The reason for this, we surmise, is that given the same number of field trials both the low and high fidelity models help inform $\theta_f$, but only the high fidelity model directly informs $\theta_h$.  When there are relatively many simulations and observations, all of the calibration parameters tend to be well constrained (panel (c)).

A subsequent simulation study is performed to compare predictions of the new model with approaches that only use some of the simulations.  Models D1 and D2 are implementations of the Kennedy and O'Hagan (2001) approach using only the low fidelity and only the high fidelity outputs, respectively.  Predictions from these models are compared with those from model D3, the proposed methodology. In other words, we are investigating whether the proposed approach of combining all simulations and observations is better in some sense than the  Kennedy and O'Hagan (2001) method using one of either the low fidelity model or high fidelity model outputs alone.

\begin{figure}[h!]
\centering
\subfigure[$n_l=40$,$n_h=10$ and $n_f=3$]{
\includegraphics[scale=0.6]{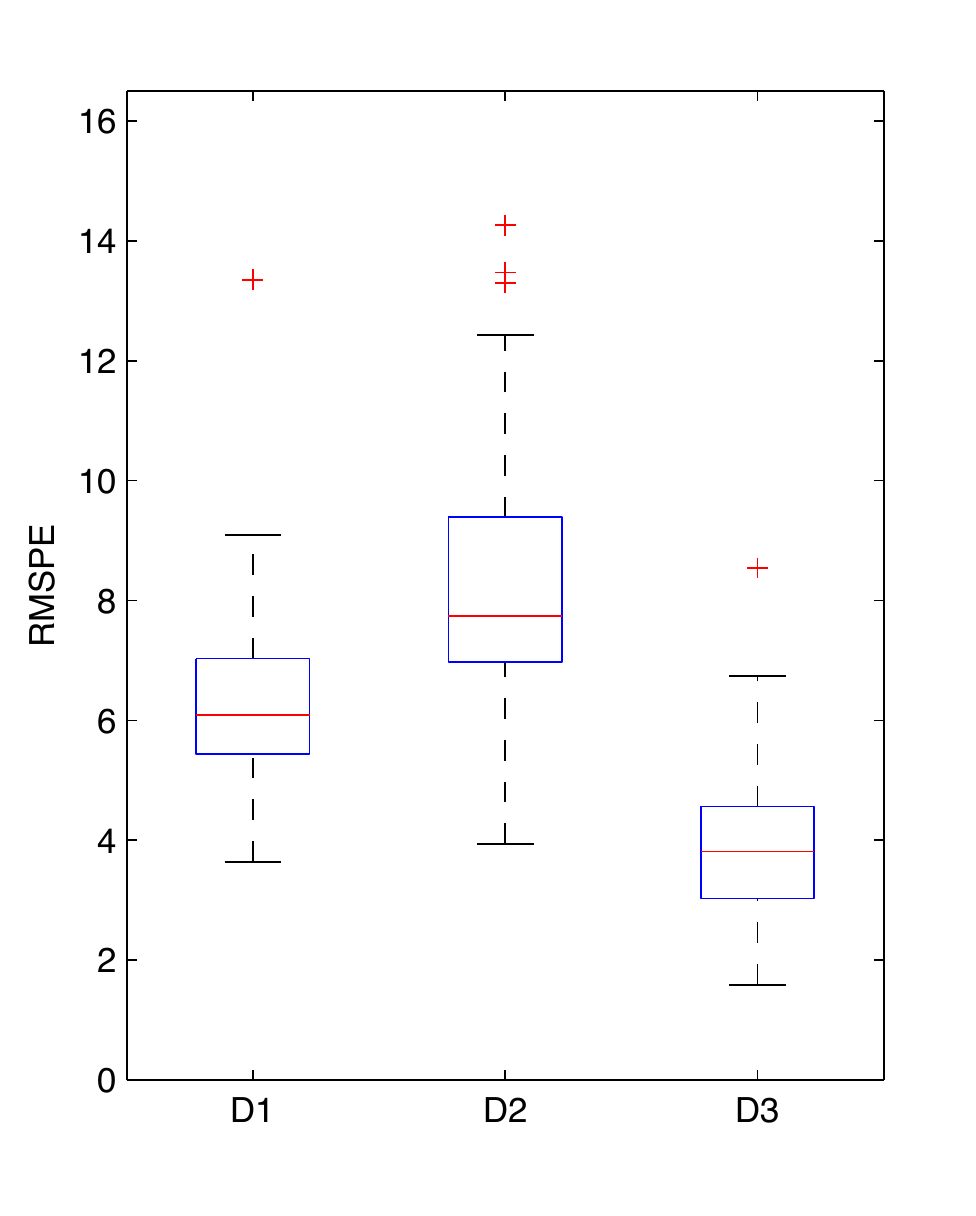}
   \label{fig:toyOrig}
 }
 \subfigure[$n_l=n_h=20$ and $n_f=3$]{
   \includegraphics[scale =0.6] {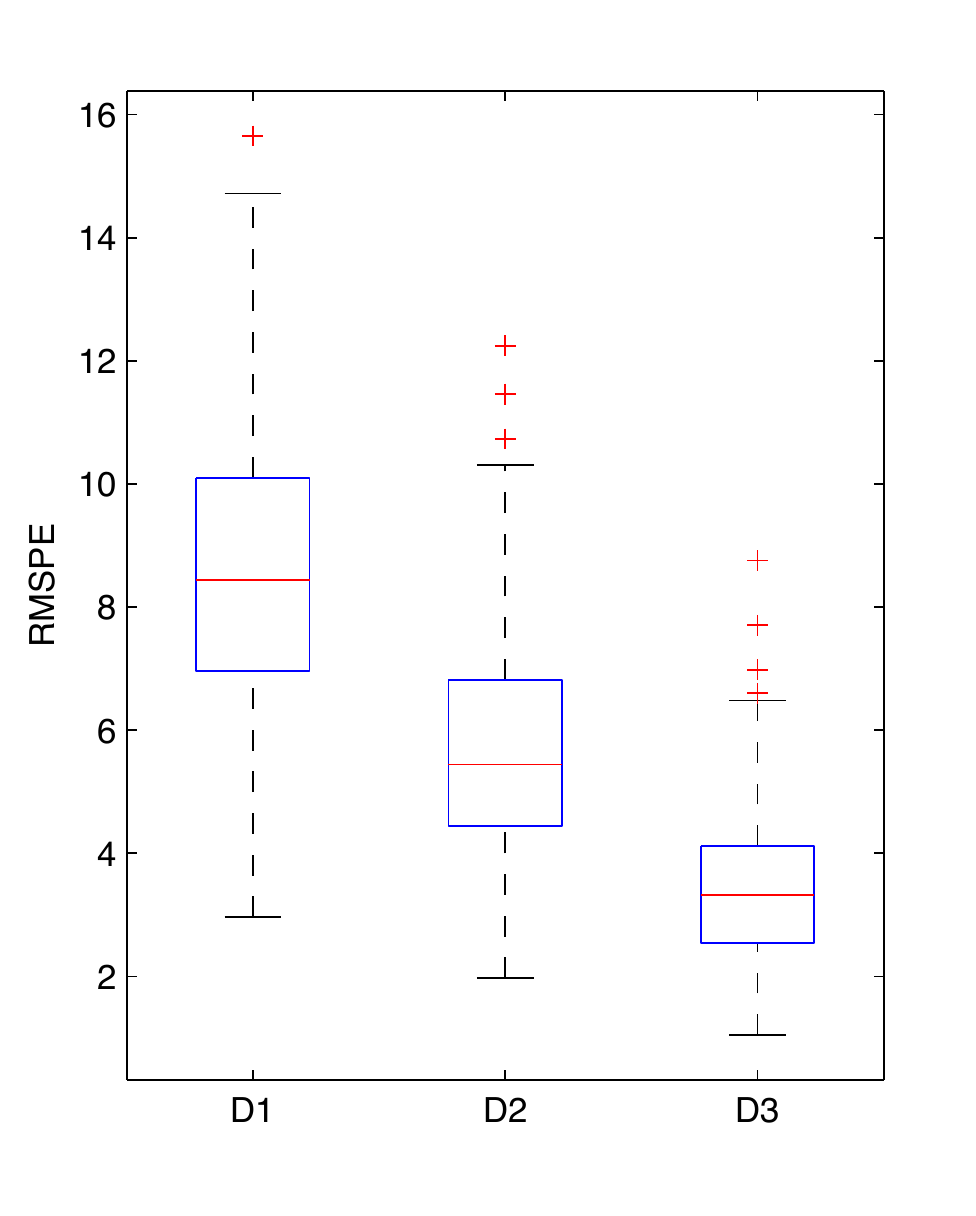}
   \label{fig:toy20Boxplot}
 }
\caption{Boxplots of the RMSPE obtained from the 100 simulated datasets analyzed using models D1-D3.}
\label{fig:toyBoxplot}\end{figure}

The simulation study is carried out as follows. Using random Latin hypercube sampling, 100 sets of training and validation data are first generated independently.  Each training set contains the same number of outputs as the above: 40 simulated values from the low fidelity simulator, 10 computer runs from the high fidelity simulator and 3 field observations.  For each simulated training set, models  D1, D2 and D3 are estimated, and  predictions of the validation set are obtained from each model.  The predictions are evaluated by computing the root mean squared prediction errors (RMSPE) for the validation data.  This is done for each of the 100 simulated training and validation datasets.  The simulation study results are summarized in Figure \ref{fig:toyBoxplot}.  

Figure \ref{fig:toyBoxplot} reveals that the RMSPE from the proposed model is consistently smaller than RMSPE of the other two models.  Interestingly, in panel (a), we notice that the RMSPE is larger for the high fidelity model than the low fidelity model.  This is the result of having relatively few runs of the high fidelity code.  Looking at Figure \ref{fig:toy20Boxplot}, when $n_l=n_h=20$ and $n_f=3$, prediction using the higher fidelity outputs does better than prediction using only the low fidelity outputs.  In either case, the proposed approach that uses all sources of data tends to do better in terms of RMSPE.

In general, we found that the proposed model that makes use of all the simulations works well in making predictions for the physical system.  The simulation demonstrates that more efficient estimation is gained through this approach.  Although calibration is not the priority, we come across a similar issue encountered by Kennedy and O'Hagan (2001) -- calibration is difficult with limited amounts of data. However, as the number of outputs increases, more information is available to calibrate the parameters of interest.  In the case of calibration, it is important to note what is being achieved.  That is, the posterior distributions reflect the uncertainty in the calibration parameters given the observations and the imperfect computer models. 

\subsection{CRASH Application}\label{sec:Crash}
\begin{figure}[h!]
\centering
\includegraphics[scale=0.7]{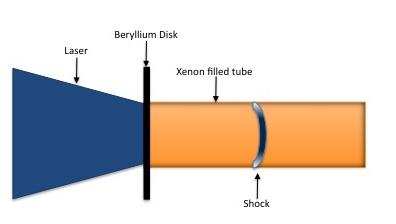}
\caption{A pictorial version of the apparatus used in the radiative shock experiments.   The black vertical bar represents the beryllium disk where the laser (in blue) deposits energy.  The shock wave breaks through the beryllium disk and moves down the xenon filled tube (horizontal bar).}
\label{fig:tube}\end{figure}

The application that motivated the proposed methodology arises from radiative shock experiments at CRASH.  Figure \ref{fig:tube} gives a diagram of the system that we want to predict. In the physical experiments, a high energy laser pulse irradiates a thin disk of beryllium at the front end of a xenon filled tube.  The energy deposited in the surface causes the beryllium to ablate.  A shock wave is then driven by the ablation pressure through the beryllium disk.  After the shock wave breaks out of the beryllium disk, the disk acts as a piston, propagating the shock at a high speed into the xenon. When the xenon is shocked, it is heated to temperatures well over 100,000 $^{o}$K and emits thermal x-ray radiation. 
These shocks are considered radiative when the radiation energy flux from the shock is high enough to impact the structure of the shock wave.  Details regarding the radiative shock physics can be found in Drake et al. (2011).  
The radiating shock experiments that we are concerned with can be viewed as small-scale experiments
for understanding astrophysical shock waves and other high temperature phenomena (McClarren et al., 2011; Drake et al., 2011).   Several measurements of interest are taken from each shock experiment and also simulations.  We focus here on the time taken for the shock wave to exit the beryllium disk (breakout time).

Using two different radiation-hydrodynamics codes (1D-CRASH and 2D-CRASH), we aim to predict the shock breakout time.  The 2D-CRASH code includes two-dimensional processes and interactions that the one-dimensional code, or 1D-CRASH, does not. As a result, the 2D-CRASH model is assumed to be able to model the experiments better than the 1D-CRASH code, but it is also more computationally expensive.  

The design variables 
 for this experiment are the thickness of the beryllium disk ($x_1$) and laser energy ($x_2$).  The electron flux limiter is calibration input to both computer models and is denoted as $t_f$.  The laser energy scale factor is an additional calibration parameter, $t_l$, required by the 1D-CRASH code but not the 2D-CRASH simulator.    The high fidelity computer code has two calibration inputs -- beryllium Gamma ($t_{h,1}$) and wall opacity scale factor ($t_{h,2}$).  All the inputs are scaled to the unit interval before fitting the data to the proposed model. 

We have 365 simulations from 1D-CRASH and 104 2D-CRASH runs available.  The designs for each computer experiment were  Latin hypercube designs, optimized using a space-filling criterion (Johnson et al., 1990). There are also 8 experiments that were conducted at the OMEGA Laser Facility at the University of Rochester where the breakout time was recorded (Boehly \etal, 1997).

The MCMC was set up as in the previous examples, with one exception.    From previous usage of the laser, it was known that the variance of the observation error was about $50\times 10^{-12}$ seconds - or approximately $1$ after standardizing.  A Gamma distribution with shape and scale parameter $(10,000, 10,000)$ was chosen for the prior of $\lambda_y$.  This is an informative prior that tightly centers the Gamma distribution at 1.  
The widths for the Metropolis updates are chosen as outlined in Section \ref{sec:hyperPrior}.  We found that convergence was achieved shortly after 1,000 MCMC steps.  So, the MCMC was run for 10,000 steps and the first 2,000 were discarded as burn-in.

Similar to the previous example, the deviations of the predictions from the observed breakout times are plotted against the predictions and the two input settings (diagnostic plots not shown). No obvious pattern is found in any of the diagnostic plots, thereby suggesting that the model fit is adequate.

\begin{figure}[h!]
\centering
\includegraphics[scale=0.7]{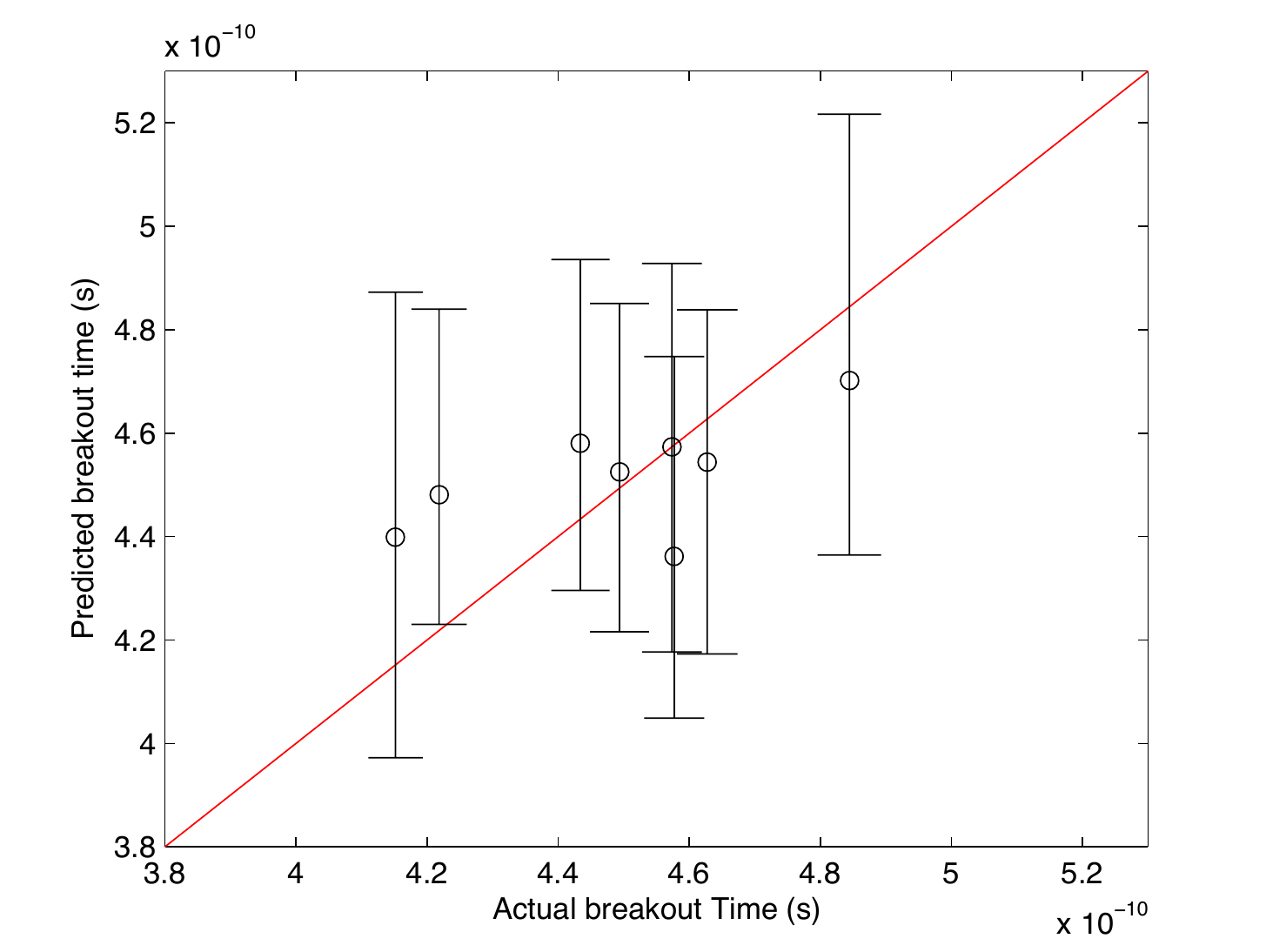}
\caption{Predicted versus actual breakout times and 95\% prediction intervals. 
}
\label{fig:predLOO}\end{figure}

A leave-one-out study is conducted to evaluate the predictive ability of the new approach.  That is, we delete an observation, fit the proposed model and predict the deleted observation.   This is done for each of the 8 observations.  Figure \ref{fig:predLOO} is a plot of the resulting predictions against the observed breakout time.  The 95\% posterior prediction interval for each point is shown in the figure.  The predictions are fairly close to the observed values and, thus most points are near to the $y=x$ line.  However, the second observation from the left gives a prediction interval that fails to capture the observation. 

\begin{figure}[h!]
\centering
\includegraphics[scale=0.6]{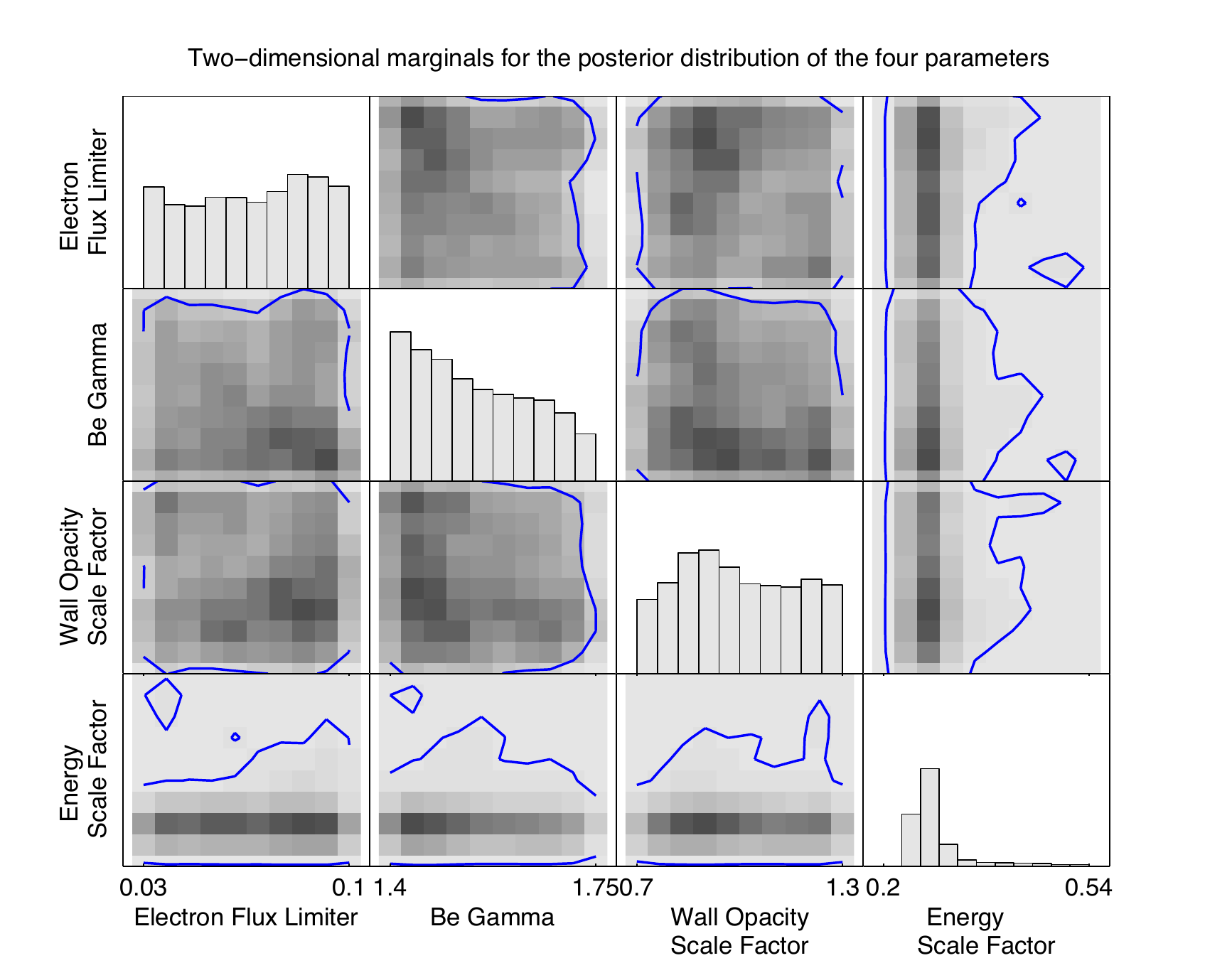}
\caption{The diagonals show the marginal posterior distributions of the calibration parameters.  The off-diagonals sub-plots contain the two-dimensional marginal posterior distributions for the four calibration parameters.  The solid lines represent the 95\% high posterior density region.
}
\label{fig:ThetaHist}\end{figure}

$\;$

Plots of the marginal posterior distributions of the calibration parameters are shown in Figure \ref{fig:ThetaHist}.  The posterior distributions for all the calibration parameters, except the energy scale factor, are not constrained in this application.  However, the posterior distributions have clear modes to suggest plausible values for the calibration parameters.

\section{Discussion}\label{sec:discussion}

So far, we have focused on the setting where there are only two computer models.  The new methodology, however, can easily be extended to model applications that involve more than two simulators.  Here, such extensions, as well as limitations of this model, are mentioned.

Suppose that there are $H$ simulators denoted as $\eta_k\left(\cdot\right) $ for $k=1,\ldots ,K $, where $\eta_k (\cdot)$ is the next highest level of fidelity model from $\eta_{k-1} (\cdot)$.  The simulators share the same design variables, $\xVec$, and some common calibration parameters, $\tVec_f$.  The additional calibration parameters required by each of the respective computer model are denoted as $\tVec_k $, for $k=1,\ldots ,K $.  The outputs of the lowest fidelity computer model are denoted as:
\[
\  Y_1\left(\xVec,\tVec_f,\tVec_1\right) = \eta_{1}\left(\xVec,\tVec_f,\tVec_1\right).
\]

The outputs from the higher fidelity simulators can then be written as a combination of the lowest fidelity simulator and discrepancy functions that capture the systematic differences between each pair of simulators. For $k=2,\ldots,K$, the simulated outputs are written:
\begin{eqnarray*}
\  Y_k\left(\xVec,\tVec_f,\tVec_h\right) &=& \eta_{k}\left(\xVec,\tVec_f,\tVec_k\right)\\
 &=& \eta_{1}\left(\xVec,\tVec_f,\thetaVec_1\right) + \mysum_{j=2}^{k-1}\delta_j\left(\xVec,\tVec_f,\thetaVec_j\right) +\delta_k\left(\xVec,\tVec_f,\tVec_k\right) .
\end{eqnarray*}

Field measurements are also available to build the predictive model.  The experimental observations are represented with the highest fidelity simulator and are written as the sum of the low fidelity simulator and discrepancy functions:
\begin{eqnarray*}
\  Y_f\left(\xVec\right) &=& \eta_{K}\left(\xVec,\thetaVec_f,\thetaVec_K\right)+\delta_f\left(\xVec\right)+\epsilon\\
 &=& \eta_{1}\left(\xVec,\thetaVec_f,\thetaVec_1\right) + \mysum_{j=2}^{K}\delta_j\left(\xVec,\thetaVec_f,\thetaVec_j\right) +\delta_f\left(\xVec\right)+\epsilon,
\end{eqnarray*}
\noindent where $\delta_f\left(\xVec\right)$ measures the discrepancy between the highest fidelity computer model and physical process.
The response surfaces of the different sources of data are modelled with GPs with mean and covariance functions discussed in Section \ref{sec:GPprior}.  

Some care should be taken in the prior specification for the precision parameters for the GPs.  We have found that the default choices of prior distributions outlined in Section \ref{sec:GPprior} work fine in most cases (e.g., the simulations in Section \ref{sec:toyEg}).  However, for some datasets, extremely large values of $\lambda_y$ are observed.  This amounts to essentially a model with no measurement error and discrepancies that are interpolating the noise.  We noticed the phenomenon when the default priors are used for the CRASH example.  This can also happen with the model proposed by Kennedy and O'Hagan (2001).  In our case, we avoided this problem because we had a more informative prior distribution for $\lambda_y$.  Alternatively,  one can address this issue by rejecting small values of a precision parameter in the MCMC (this was done in Higdon et al. (2004)), or at the design stage by taking replicate field observations. 


A further note of caution with respect to the experimental design.  The design regions for the computer experiments should coincide to avoid uncertainty due to extrapolation in the discrepancies between models.  Suppose for example, the design for $\tVec_f$ in the low fidelity simulator explores a much larger region than the design for the high fidelity model.  When predictions are made, the proposed approach averages over the posterior distribution of the calibration parameters.  For values of $\bs\theta_f$ from the posterior that are outside of the range explored by the design of the high fidelity model, the proposed approach extrapolates $\delta_2(\cdot)$.  This results in larger prediction intervals.  

Lastly, we do not address problems with design variables that appear in only some, but not all, simulators.  This is a topic for future work.

 
\section{Conclusion}\label{sec:conclusion}
A new methodology, which combines outputs from multi-fidelity computer models and field observations, is proposed.  The approach successfully uses a Bayesian hierarchical model to make predictions of the physical system with associated measurements of uncertainty (e.g., posterior variance or  prediction intervals).   Different GPs are used to model the various response surfaces.  The real example that motivated this work used two simulators of the process, but methodology can be easily extended to cases with more than two simulators.

\newpage
\renewcommand{\baselinestretch}{1.0}
\centerline{\large {\bf References}}

\def\beginref{\begingroup
                \clubpenalty=10000
                \widowpenalty=10000
                \normalbaselines\parindent 0pt
                \parskip.0\baselineskip
                \everypar{\hangindent1em}}
\def\endref{\par\endgroup}
\beginref

Bartos, L. S. and O'Hagan, A. (2009). ``Diagnostics for Gaussian Process Emulators''. {\em Technometrics}, {\bf 51}, 425-438.

Bayarri, M.J., Berger, J. O., Paulo, R., Sacks, J., Cafeo, J. , Cavendish, J., Lin, C. and Tu, J. (2007).``A Framework for Validation of Computer Models''. {\em Technometrics}, {\bf 49}, 138-154.

Boehly, T. R., Brown, D. L., Craxton, R. S., Keck, R. L., Knauer, J. P., Kelly, J. H., Kessler, T. J., Kumpan, S. A., Loucks, S. J., Letzring,  S. A., Marshall, F. J., McCrory, R. L., Morse, S. F. B., Seka, W., Soures, J. M. and Verdon, C. P. (1997). ``Initial performance results of the OMEGA laser system'' {\em Optics Communications}, {\bf 133}, 495-506.

Craig, P. S., Goldstein, M., Rougier, J. and Seheult, A. H. (1998). ``Bayesian Forecasting for Complex Systems Using Computer Simulators''. {\em The Statistician}, {\bf 47}, 37-53.

Craig, P. S., Goldstein, M., Seheult, A. H. and Smith, J. A. (2001). ``Constructing Partial Priors Specifications for Models of Complex Physics Systems''. {\em Journal of the American Statistical Association}, {\bf 96}, 717-729.

Cumming, J. A. and Goldstein, M. (2009). ``Small Sample Bayesian Designs for Complex High-Dimensional Models Based on Information Gained Using Fast Approximations''. {\em Technometrics}, {\bf 51}, 377-388.

Drake, R. P., Doss, F. W., McClarren, R. G., Adams, M. L., Amato, N., Bingham, D., Chou,  C. C., DiStefano, C., Fidkowski, K., Fryxell, B., Gombosi, T. I., Grosskopf, M. J., Holloway, J. P., van der Holst, B., Huntington, C. M., Karni, S., Krauland, C. M., Kuranz, C. C., Larsen, E.,van Leer,  B., Mallick, B., Marion, D., Martin, W., Morel, J. E., Myra, E. S., Nair, V., Powell, K. G., Rauchwerger, L., Roe, P., Rutter, E., Sokolov, I. V., Stout, Q., Torralva, B. R., Toth, G., Thornton, K. and Visco, A. J. (2011). ``Radiative effects in radiative shocks in shock tubes''. {\em High Energy Density Physics}, {\bf 7}, 130-140.

Graves, T. (2005). ``Automatic Step Size Selection in Random Walk Metropolis Algorithm''. Technical report, Los Alamos National Laboratory.

Gelman, A., Carlin, J. B., Stern, H. S. and Dublin, D. B. (2004). ``Bayesian Data Analysis'' {\em 2nd edition.} Chapman \& Hall, Boca Raton.

Goldstein, M. and Rougier, J. (2006). ``Bayes Linear Calibrated Prediction for Complex Systems''. {\em Journal of the American Statistical Association}, {\bf 101}, 1132-1143.

Hastings, W. K. (1970). ``Monte Carlo Sampling Methods Using Markov Chains and Their Aplications''. {\em Biometrika}, {\bf 57}, 97-109.

Higdon, D., Kennedy, M., Cavendish, J., Cafeo, J. and Ryne, R. D. (2004). ``Combining Field Data and Computer Simulations for Calibration and Prediction". {\em SIAM Journal of Scientific Computing}, {\bf 26}, 448-466.

Higdon, D., Gattiker, J., Williams, B. and Rightley, M. (2008). ``Combining Field Data and Computer Simulations for Calibration and Prediction". {\em Journal of the American Statistical Association}, {\bf 103}, 570-583.

Johnson, M.E., Moore, L.M., Ylvisaker, D. (1990), ``Minimax and maximin distance designs". {\em Journal of
Statistical Planning and Inference}, {\bf 26}, 131-148.

Linkletter, C., Bingham, D., Hengartner, N., Higdon, D. and Ye, K.Q. (2006). ``Variable Selection for Gaussian Process Models in Computer Experiments''. {\em Technometrics}, {\bf 48}, 478-490.

McClarren, R.G., Ryub, D., Drake, P., Grosskopf, M., Bingham, D., Chou, C-C., Fryxell, B., van der Holst, B., Holloway, J.P., Kuranz, C.C., Mallick, B., Rutter, E. and Torralva, B. (2011).  ``A Physics Informed Emulator for Laser-Driven Radiating Shock Simulations�. {\em Reliability Engineering and System Safety}, {\bf 96}, 1194-1207.

Metropolis, N., Rosenbluth, A., Rosenbluth, M., Teller, A., and Teller, E. (1953). ``Equations of State Calculations by Fast Computing Machines". {\em Journal of Chemical Physics}, {\bf 21}, 1087-1091.

Oakley, J. E. and O'Hagan, A. (2004). ``Probabilistics Sensitivity Analysis of Complex Models: A Bayesian Approach''. {\em Journal of the Royal Statistical Society, Ser. B}, {\bf 66}, 751-769.

Qian, Z. G., Seepersad, C. C., Joseph, V.R., Allen, J.K. and Wu, J. C. F. (2006). ``Building Surrogate Models Based on Detailed and Approximate Simulations''. {\em ASME Journal of Mechanical Design}, {\bf 128}, 668-677.

Qian, P. Z. G. and Wu, J. C. F. (2008). ``Bayesian Hierachical Modeling for Integrating Low-Accuracy and High-Accuracy Experiments''. {\em Technometrics}, {\bf 50}, 192-204.

Reese, C. S., Wilson, A. G., Hamada, M., Martz, H. F. and Ryah, K. J. (2004). ``Integrated Analysis of Computer and Physical Experiments''. {\em Technometrics}, {\bf 46}, 153-164.

Rougier, J., Sexton, D. M. H., Murphy, J. M. and Stainforth, D. (2009). ``Analyzing the Climate Sensitivity of the HadSM3 Climate Model Using Ensembles From Different but Related Experiments''. {\em Journal of Climate}, {\bf 22}, 3540-3557.

Sacks, J., Welch, W. J., Mitchell, T. J. and Wynn, H. P. (1989). ``Design and Analysis of Computer Experiments''. {\em Statistical Science}, {\bf 4}, 409-423.

Santner, T. J., Williams, B. J. and Notz, W. I. (2003). ``The Design and Analysis of Computer Experiments''. {\em New York: Springer-Verlag}.

Welch, W. J., Buck, R. J., Sacks, J., Wynn, H. P., Mitchell, T. J. and Morris, M. D. (1992). ``Screening, Predicting, and Computer Experiments''. {\em Technometrics}, {\bf 34}, 15-25.

Williams, B., Higdon, D., Gattiker, J., Moore, L., McKay, M. and Keller-McNulty, S. (2006). ``Combining Experimental Data and Computer Simulations, With an Application to Flyer Plate Experiments". {\em Bayesian Analysis}, {\bf 1}, 765-792.

\endref
\end{document}